\documentclass[12pt]{article}
\usepackage{amsmath}
\usepackage{graphicx}
\usepackage{enumerate}
\usepackage{natbib}
\usepackage{url} 

\newcommand{\blind}{1}

\begin{document}

\def\spacingset#1{\renewcommand{\baselinestretch}%
{#1}\small\normalsize} \spacingset{1}


\if1\blind
{
  \title{\bf Discussion of ``A Gibbs sampler for a class of random convex polytopes''}
  \author{Jonathan P Williams \\
    Department of Statistics, North Carolina State University}
  \maketitle
} \fi

\if0\blind
{
  \bigskip
  \bigskip
  \bigskip
  \begin{center}
    {\LARGE\bf Discussion of ``A Gibbs sampler for a class of random convex polytopes''}
\end{center}
  \medskip
} \fi

\bigskip
\begin{abstract}
An exciting new algorithmic breakthrough has been advanced for how to carry out inferences in a Dempster-Shafer (DS) formulation of a categorical data generating model.  The developed sampling mechanism, which draws on theory for directed graphs, is a clever and remarkable achievement, as this has been an open problem for many decades.  In this discussion, I comment on important contributions, central questions, and prevailing matters of the article.
\end{abstract}

\vfill

\section{Introduction}

I commend the authors for their efforts in revisiting the DS framework for statistical inference, but from a computational perspective.  It is perhaps untraditional for the Theory and Methods section of the Journal of the American Statistical Association to publish a manuscript with contributions that are without methodological innovation, but purely computational in nature.  This speaks to the importance of the problem the authors address; it seems that this has been an open problem for many decades.

I begin my comment piece with a reiteration of the authors' concluding remark that the field of statistics lacks a unifying foundation for parametric inference, and I add, for how to understand data more broadly.  Much of foundations research is thought of as a novelty by the broader statistics community; philosophical discussions about how to think about data are dismissed as  inconsequential and peripheral to contemporary data problems.  This has largely been driven in modern decades by efforts to chase the algorithmic development success of more computationally oriented experts in the mathematical and computer sciences in their efforts to solve engineering data problems/tasks.  

The significance of foundations questions have also been obscured by an obsession, in statistics communities, with appeals to asymptotic justification for methods research, and a lack of emphasis for studying finite-sample properties of new methods.  Arguably, the distractions from addressing more foundational questions in statistical inference is in large part because these types of investigations do not lend themselves to be packaged as succinct or affirming mathematical statements/characterizations.  For instance, how to analyze a model/variable selection method in the absence of a sparsity assumption?  Surely, sparsity is {\em not} an assumption that is believed to be true as often as it is taken as an assumption in academic papers.  This is an issue that I have attempted to address in the series of articles, \cite{williams2019,williams2019a}, as well as in ongoing work.  

Nonetheless, rather than developing an argument, here, for the relevance and importance of foundations research, in what follows are comments organized into sections focused on important contributions, central questions, and prevailing matters of the article under discussion.

\section{Important contributions}

The genius of the article under discussion is the insightful connection made between the expression for the data generating, or structural equation of a categorical random variable,
\begin{equation}\label{dge}
x_{n} = \sum_{k\in[K]}k\mathbf{1}\{u_{n}\in \Delta_{k}(\theta)\},
\end{equation}
and the framework of directed graphs.  This connection is explicitly characterized by the authors' Proposition 3.2, where, conditional on a sample of auxiliary points, $u_{1},\dots,u_{N}$, a closed-form expression of some $\theta \in \Delta$ is constructed that is consistent with the observed data in equation (\ref{dge}).  That is, the $\theta$ value constructed via Proposition 3.2 is contained in the set,
\[
\mathcal{F}(\mathbf{u}) := \{\theta \in \Delta : \forall n \in [N], u_{n} \in \Delta_{x_{n}}(\theta)\},
\]
so that the set $\mathcal{R}_{\text{x}} := \{\mathbf{u} : \mathcal{F}(\mathbf{u}) \ne \emptyset\}$ is guaranteed to be non-empty.  The elegantly constructed $\theta$ in Proposition 3.2 has components proportional to the inverse of the exponentiated directed graph path ``value'', minimized over ratios of components $\frac{u_{n,\ell}}{u_{n,k}}$ over all $n \in \{1,\dots,N\}$.  

Thus, given a sample of auxiliary points, $u_{1},\dots,u_{N}$, the authors provide a mechanism to update $\theta$ consistent with the data generating equation (\ref{dge}) so that it is possible to uniformly sample a new set of auxiliary points via an original algorithm proposed in \cite{dempster1966} (this algorithm is re-stated as Algorithm 1 in the article under discussion).  The contribution is summarized as Algorithm 2 in the article under discussion.

Algorithm 1, re-stated from \cite{dempster1966}, provided a very simple mechanism for sampling a value $u$ on the sub-simplex $\Delta_{k}(\theta)$, but did not resolve the issue of updating $\theta$.  So it seems that the article under discussion has resolved this critical link to construct a sampling mechanism for uniform sampling of the auxiliary components of the data generating equation (\ref{dge}).  Certainly this is a remarkable accomplishment.

Second, the authors do a great job in the article of highlighting the role of the structural equation in a fiducial approach to inference in the categorical model.  In particular, the mapping $\mathcal{F} : \mathbf{u} \to \theta$ is the inverse mapping of the data generating equation (\ref{dge}).  Such a notion/construction characterizes a whole class of contemporary fiducial approaches.  Two notable examples are the inferential models (IM) approach \citep{martin2015c} and the generalized fiducial inference (GFI) approach \citep{hannig2016}.  The IM approach is especially relevant because in the IM framework belief and plausibility functions also drive the statistical inference, where questions of interest are formulated as measurable sets in the parameter space.  Accordingly, just as described for the DS approach, the IM approach admits a ``don't know'' probability, denoted as $r$ in the belief, plausibility, and ``don't know'' triple, $(p,q,r)$, respectively.  

The importance of this third, ``don't know'' category of inference has been drawing attention in recent years, driven in part by the discovery of the {\em false confidence theorem} introduced in \cite{balch2019}.  The consequence of the false confidence theorem is illustrated in \cite{balch2019} in the context of a phenomenon observed in a Bayesian analysis to assess the probability of satellite collision based on real satellite trajectory data.  It is observed that a degradation of the data quality leads to a decreased posterior probability of a collision event.  Obviously this is not because the analyst can be {\em more} sure of non-collision, but the phenomenon ensues because the analyst is necessarily {\em less} sure of a collision event.  Under the usual axioms of probability calculus, probabilities of mutually exclusive but collectively exhaustive events must sum to one.  Since the logic for Bayesian inference is tied to the estimation of a posterior probability distribution, this problem is foundational for Bayesian theory.  In my own work, namely \cite{carmichael2018}, we provide an illustration of simple examples where/how the false confidence theorem manifests.

An aspect that the article under discussion lacks is a more complete literature review and comparison relating the DS approach to other fiducial approaches, most importantly because the value of the contribution of the article resides in the relevance of the DS approach in the context of the modern literature on fiducial ideas.  For instance, \cite{martin2021} showed that, in general, valid and efficient tests and confidence regions are fundamentally tied to the distributions of random subsets on the parameter space.  Moreover, the IM approach, which parallels (at least this formulation of) inference in the DS framework, has exhibited growing success over the past decade, both in terms of methodological developments and computational strategies for implementation (see, for example, \cite{martin2015a,martin2015b,martin2015c} as well as \cite{martin2012,martin2018,cella2019,martin2019}).  Nonetheless, other applications of IM may have non-trivial non-emptiness constraints for the distribution of random sets of the auxiliary variable.  These applications, similar to the article being discussed, may require more sophisticated computational strategies.  Accordingly, it can be expect that the authors' ideas and developments will have an impact far beyond multinomial applications, to valid and efficient probabilistic inference more generally.  

Next, the GFI approach also begins with a data generating equation, say $Y = G(\gamma, U)$, for some deterministic function $G$, unknown parameter $\gamma$, and auxiliary random variable $U$ with some known distribution that is independent of $\gamma$.  While DS and IM are approaches that infer uncertainty about the unknown parameter $\gamma$ by constructing random sets that capture the variability in $U$, the GFI approach is to employ a ``switching principle'' whereby a distributional estimator of the unknown parameter $\gamma$ is inherited from the distributional assumption on auxiliary random variable $U$.  From the perspective of DS ideas, an argument can be made that the switching principle is philosophically problematic, but nonetheless, it has been demonstrated that GFI leads to statistically principled reasoning for inference \citep{hannig2016,liu2016}.  Moreover, it continues to be demonstrated in the literature that GFI approaches to contemporary data problems lend themselves to viable computational strategies (see, for example, \cite{hannig2014,lai2015,williams2019, williams2019a,cui2019,li2020,williams2020,shi2021}).  

It is also important to afford reference to the growing literature on confidence distributions (e.g., \cite{xie2013,schweder2016}).  Successful development of computational strategies in the DS framework are exciting and encouraging, but they should not be evaluated in a vacuum.

\section{Central questions}

The article under discussion represents an important contribution to the contemporary literature, but of course, much more work is needed to establish the DS framework as more than a novelty.  The central questions for discussion that arise from the article are the following:
\begin{enumerate}
\item Is the DS approach worth revisiting?
\item Does the computational strategy developed in the article represent a viable path forward for real data analyses within a DS framework?
\end{enumerate}

Certainly an affirming answer to the second question would provide a positive answer to the first.  With this thought in mind, the most apparently limiting aspect of the proposed algorithm for the DS inference exclusively constructed for the categorical data model, is the feature that the auxiliary random variables $U_{1}, \dots, U_{N}$ are elements of the same ambient space as the parameter $\theta$.  This is problematic because the authors' entire algorithmic development intrinsically exploits this fact.  The exposition of the development of the ideas is a series of geometric arguments that directly relate ratios of components of $\theta$ to ratios of the values of components of the $u_{1}, \dots, u_{N}$ via $\eta_{k\to\ell}(u_{n})$.  It is not obvious at all how these ideas could be extended to apply more generally for a broader class of data generating, or structural equations.

A question of secondary concern is what to do if the parameters in a given data generating equation are not supported on a simplex?  While there is only so much responsibility that any one article can be charged with, this is an important question for the computational feasibility of the DS approach, more generally.  Moreover, since this is a discussion article with such a central theme, this question stands out as relevant.  The authors state in their introduction that over past decades, DS theory has seen ``various applications in signal processing, computer vision and machine learning \dots.''  It would be helpful if the authors could provide an overview of how DS computations are carried out in the referenced application areas.  What makes these other settings more amenable to DS computations than the categorical model?

\section{Prevailing matters}

Section 4.2 in the article under discussion, as it reads, is a bit of a hard sell for why, as the authors state in their concluding remarks: ``One of the appeals of the DS framework is its flexibility to incorporate types of partial information which are difficult to express in the Bayesian framework.''  This ties into the broader theme of the article about whether DS is worth considering by mainstream statistical communities.  Arguably, the overwhelming majority of practitioners that incorporate prior information in a typical Bayesian inference fashion would find Dempster's rule of combination an inaccessible obfuscation of how to formalize prior information in a statistical inference.  Surely, this is an ill-founded a reason to dismiss an important idea, but it {\em does} represent a real obstacle for transforming the DS approach into a class of tools that are readily accessible by a wide audience of practitioners.

I hope that in future work, the authors will continue to develop on other equally impressive ideas that increase the computational feasibility of the DS approach.  Furthermore, it would be very exciting to see an investigation of the comparison of the DS approach to a modern Bayesian approach, on a real data set of practical significance.  A Bayesian analysis would be very problematic if the excluded ``don't know'' probability has a non-trivial value, as demonstrated for satellite trajectory data analyses in \cite{balch2019}.  In theory, the DS approach will not suffer from the inadequacy exhibited in the Bayesian approach.

\bibliographystyle{agsm}
\bibliography{references}
\end{document}